\documentclass[preprint]{aastex}

\usepackage{amssymb,amsmath,amsbsy}
\usepackage{booktabs}
\usepackage[caption=false]{subfig}

\newcommand{\project}[1]{\textsl{#1}}
\newcommand{\gaia}{\project{Gaia}}
\newcommand{\spitzer}{\project{Spitzer}~}
\newcommand{\given}{\,|\,}

\newcommand{\msun}{\mathrm{M}_\odot}

\newcommand{\D}{{\bf D}}
\newcommand{\W}{{\bf W}}

\newcommand{\J}{{\boldsymbol J}}
\newcommand{\bSigma}{{\bf \Sigma}}

\newcommand{\rtide}{r_{\rm tide}}
\newcommand{\bs}{\boldsymbol}
\newcommand{\sat}{{\rm p}}
\newcommand{\tub}{t_{\rm ub}}

\newcommand{\tailbit}{\beta}
\newcommand{\Loffset}{\alpha}

\newcommand{\vhalo}{v_{\rm h}}
\newcommand{\rewinder}{\texttt{Rewinder}}
\newcommand{\potp}{$q_1,q_z,\phi,\vhalo$}

\begin{document}

\title{Inferring the gravitational potential of the Milky Way with a few precisely measured stars}
\author{Adrian M. Price-Whelan\altaffilmark{\colum,\adrn}, 
	    David W. Hogg\altaffilmark{\nyu,\cds,\mpia}, 
	    Kathryn V. Johnston\altaffilmark{\colum}, 
	    David Hendel\altaffilmark{\colum}}

\newcommand{\colum}{1}
\newcommand{\adrn}{2}
\newcommand{\nyu}{3}
\newcommand{\cds}{4}
\newcommand{\mpia}{5}
\altaffiltext{\colum}{Department of Astronomy, 
		              Columbia University, 
		              550 W 120th St., 
		              New York, NY 10027, USA}
\altaffiltext{\adrn}{To whom correspondence should be addressed: adrn@astro.columbia.edu}
\altaffiltext{\nyu}{Center for Cosmology and Particle Physics,
                      Department of Physics, New York University,
                      4 Washington Place, New York, NY, 10003, USA}
\altaffiltext{\cds}{Center for Data Science,
                      New York University,
                      4 Washington Place, New York, NY, 10003, USA}
\altaffiltext{\mpia}{Max-Planck-Institut f\"ur Astronomie,
                     K\"onigstuhl 17, D-69117 Heidelberg, Germany}

\begin{abstract}
The dark matter halo of the Milky Way is expected to be triaxial and filled with substructure. It is hoped that streams or shells of stars produced by tidal disruption of stellar systems will provide precise measures of the gravitational potential to test these predictions. We develop a method for inferring the Galactic potential with tidal streams based on the idea that the stream stars were once close in phase space. Our method can flexibly adapt to any form for the Galactic potential: it works in phase-space rather than action-space and hence relies neither on our ability to derive actions nor on the integrability of the potential. Our model is probabilistic, with a likelihood function and priors on the parameters. The method can properly account for finite observational uncertainties and missing data dimensions. We test our method on synthetic datasets generated from N-body simulations of satellite disruption in a static, multi-component Milky Way including a triaxial dark matter halo with observational uncertainties chosen to mimic current and near-future surveys of various stars. We find that with just four well-measured stream stars, we can infer properties of a triaxial potential with precisions of order 5-7 percent. Without proper motions we obtain 15 percent constraints on potential parameters and precisions around 25 percent for recovering missing phase-space coordinates. These results are encouraging for the eventual goal of using flexible, time-dependent potential models combined with larger data sets to unravel the detailed shape of the dark matter distribution around the Milky Way.
\end{abstract}

\keywords{
  Galaxy: kinematics and dynamics
  ---
  Galaxy: halo
  ---
  cosmology: dark matter
}

\section{Introduction}\label{sec:intro}

Early large-scale, cosmological simulations of galaxy formation in the $\Lambda$CDM paradigm suggested that the spherically-averaged density profiles of dark matter halos follow a universal profile across a large dynamic range in mass \citep{navarro96}. Since then, higher resolution simulations --- both with and without baryons --- have produced dark matter halos that (1) are permeated with substructure on many scales, (2) are triaxial in shape, and (3) have shapes and orientations that vary with radius \citep{dubinski91, jing02, kuhlen07, veraciro11}. Dark-matter-only simulations produce triaxial halos \citep{jing02} with large density fluctuations \citep{zemp09}. Inclusion of baryons tends to soften the triaxiality and graininess in the inner galaxy through a combination of dissipative infall \citep{dubinski94} or cooling \citep{bryan13}. These processes combined with the gravity from a baryonic disk or ellipsoid can act to make the inner halo more oblate or spherical, however they do not seem to erase the clumpy, triaxial nature of the outer halo \citep[e.g.,][]{pontzen12}. This can lead to radially-dependent axis ratios, orientation, and smoothness, and suggests that the true mass distributions around Milky Way-like galaxies are not easily represented by simple, time-independent potentials. Methods that seek to measure the gravitational potentials around such galaxies must be flexible enough to handle generic potential forms where finding simple analytic approximations or computing actions may not be possible.

The bulk of the baryonic matter in galaxies spans roughly 5-10\% of the spatial extent of the host dark matter halo. Hence, the brightest and most easily observable components of a galaxy are sensitive to the inner portion of the host halos mass distribution. For example, the rotation curves of disk galaxies trace the inner mass with exquisite sensitivity since matter in disks can be assumed to move on nearly circular orbits. Measurements of the dark matter distribution at large radii is complicated by the low density of visible tracers, observational difficulties of measuring kinematics of stars at large distances, and unknown orbits. Around external galaxies, the extended mass distribution has been studied using a variety of approaches \citep[see][for a a complete and detailed review]{courteau13}. For example, the kinematics of tracer populations such as globular clusters or planetary nebulae can be used to derive mass estimates under the assumptions that these satellite systems are on random orbits and are well-mixed in orbital phase \citep[early investigations include][]{mendez01,cote03}. Simple, parameterized fits to both the mass and orbit distribution have been simultaneously constrained using such data \citep[e.g.][]{napolitano11,deason12c}. Alternatively, the statistical properties of gravitationally lensed background sources around a galaxy can be used to constrain the \emph{projected} shape, orientation, and radial profile of mass \citep[see, for example, the Lens Structure and Dynamics Survey described in][]{koopmans02}. Of course, lensing reconstructions can only be performed for galaxies which closely intersect our line of sight to background sources, but the advent of large photometric catalogues has allowed automatic searches for such chance alignments and significant increases in the number of objects studied in this way \citep[e.g. the Sloan Lens ACS Survey, see][]{bolton06}.

Within the Milky Way our unique vantage point allows us a three-dimensional view of stars within our own dark matter halo. Our proximity allows us to use individual stars as kinematic tracers and hence build much larger samples that probe deeper into the halo than the globular cluster and planetary nebula studies of external galaxies. For example, \cite{deason12a} used halo BHB stars selected from the Sloan Digital Sky Survey \cite[SDSS;][]{york00} as a random tracer population to measure the mass and slope of a power-law fit to the potential. Such studies assume that the tracer orbits are randomly sampled from a smooth distribution function and are fully phase mixed. However, large photometric surveys such as the SDSS and 2MASS \citep{skrutskie06} have discovered copious amounts of substructure --- in streams and kinematic associations of stars --- in the Milky Way halo \citep[e.g.,][]{belokurov06, rochapinto04}, thus demonstrating that the stellar distribution is neither on random orbits nor fully phase-mixed. Substructure in the form of stellar streams and clouds is known to bias mass and velocity inferences from random tracer methods by several tens of percent \citep{yencho06}.

Another approach to using halo stars as potential measures is to exploit the non-random nature of halo. Tidal streams are dynamically cold systems --- debris typically have small distributions of energy and angular momentum --- and thus require orders of magnitude fewer tracers than a random sample to get constraints of comparable accuracy to Jeans analysis. For example, in the simplest case we might assume that debris stars are actually still on the same orbit as their progenitor system (a \emph{wrong} assumption, see below). This information about the orbits combined with measurements of the full-space velocities ${\bf v}$ at different points ${\bf x}$ in the structure (e.g., along a stream) would give us a direct measure of differences in a potential, $\Phi$.

\citet[][LM10]{law10} used N-body simulations of the disruption of the Sgr dwarf galaxy to simultaneously fit a model to the available data on the debris stars and a triaxial, analytic Milky Way potential. By varying parameters of the Milky Way potential, they found ``best-fit'' parameters by comparing the properties of observed Sgr stars to their simulated debris. The computational costs of running N-body simulations limited their search to a grid of potential parameters and forced the authors to fix many other parameters (e.g., properties of the disk and bulge). Nevertheless, they were able to constrain the 3D shape that their assumed potential model must take in order to best represent the Milky Way out to $\sim$70 kpc and found that the best-fitting halo has a nearly oblate (only mildly triaxial) shape flattened in a direction close to the Sun-Galactic center line. Though an unlikely orientation for the halo --- \cite{debattista13} find that the disk of the Milky Way would not remain stable in such a configuration --- LM10 showed that the data is at a state where such inference is possible. 

The computational costs associated with N-body simulations has motivated the development of many methods that approximately model tidal streams. The simplest alternative is to fit a single orbit to observed debris \citep[e.g.,][]{koposov10, deg13}. Though this is known to be incorrect and leads to biases in inferred properties of the underlying potential \citep[e.g.,][]{eyre11, lux13, sanders13a}, \cite{deg14} and \cite{lux13} have used orbit fitting to demonstrate the power of combining multiple streams in dynamical inference. To account for the offset between the orbit of the progenitor and the orbits of the debris stars, methods have been proposed that add some dispersion or offset around a single orbit either in phase space \citep[e.g.,][]{eyre09a, varghese11, kuepper12} or action-angle coordinates \citep{eyre11, sanders13b, bovy14, sanders14}. Other statistical methods have been proposed \citep[][]{johnston99a, penarrubia12, sanderson14} that may prove powerful when applied to, for example, data from the \gaia\, mission, where full 6D coordinates will be known for large samples of stars in the halo --- and therefore many debris structures --- but stream membership is not known for all stars.

Motivated by the strengths of previous work, we identify a minimum set of issues that any approximate stream model or potential recovery method should address (see Section~\ref{sec:discussion} for a more detailed discussion of these points in the context of this work):
\begin{enumerate}
	\item \textbf{Observational uncertainties:} The known debris structures are $\gtrsim$10 kpc from the Sun where distance and proper motion measurement errors are significant. Thus it is critical for any method that uses tidal debris to incorporate observational uncertainties and missing dimensions in a consistent and justified way. 
	\item \textbf{Form of the potential:} There is large uncertainty in the radial profile, shape, orientation, and graininess of the outer halo and the constancy of these parameters over distance. Properly describing the potential may require non-parametric techniques or complex analytic functions so the method should not rely on the existence of or ability to compute conserved orbital properties such as actions.
	\item \textbf{Multiple debris structures:} Near-future photometric surveys such as \gaia\, and the \project{LSST} will likely discover many new streams and kinematic associations of stars. Potential recovery methods should be able to simultaneously use multiple streams and incorporate other dynamical constraints.
	\item \textbf{Comparing models to data:} Matching generated streams to observed stellar densities is difficult; models that rely on this must account for observational biases, internal properties of the progenitor, and background halo stars. 
	\item \textbf{Computational expense:} Full N-body simulations are expensive to run; incorporating an N-body simulation into a likelihood function evaluation and then performing a parameter search would be computationally intensive and, presently, intractable.
\end{enumerate}

In \citet{apw13}, we introduced a simple method based on the work of \citet{johnston99a} for using individual stars associated with tidal debris combined with knowledge about the mass and orbit of the progenitor to constrain properties of the host galaxy potential. The method exploits the relationship between the phase-space distribution of debris and (measurable) properties of the progenitor system (e.g., the tidal radius and escape velocity).  Specifically, a better potential is one in which orbits of test particle stars (integrated backwards from their present position) came close (within the tidal radius in position, and escape velocity in velocity) to the orbit of the progenitor at some time in the past. 

In this paper, we present a fully probabilistic model (\rewinder) for tidal streams that builds on these simple scalings, which depend only on the mass and orbit of the progenitor and parent potential. By ``probabilistic model'' we mean a justified, parametrized likelihood function with priors on the parameters. \rewinder\ relies on numerical orbit integration in ordinary phase-space and can thus incorporate arbitrarily complex forms for the parent potential (such as time dependence and significant substructure). Individual stars are constraints on the potential, thus \rewinder\ can handle very small samples of well-measured stars (e.g., 4-16). Incorporating multiple streams, debris structures, and other kinematic information is trivial but will be explored in future work.

In Section~\ref{sec:sims} we describe a suite of N-body simulations that span a range of progenitor masses on a characteristic, mildly eccentric orbit and then use them in Section~\ref{sec:method} to motivate a new, flexible model for tidal streams that works entirely in phase-space. In Section~\ref{sec:experiments} we demonstrate how \rewinder\ can be used to measure properties of a non-trivial Galactic potential by performing several experiments with simulated observations of data from N-body simulations. In Section~\ref{sec:discussion}, we discuss the results of these experiments and the extent to which we address the above points 1-5. We conclude in Section~\ref{sec:conclusion}.

\section{Simulations}\label{sec:sims}
We performed a set of N-body simulations with the Self-Consistent Field (SCF) basis function expansion code \citep{hernquist92} to build realistic models of streams for studying the phase-space distribution of debris as it is stripped from its progenitor. We later use one of these simulations to test \rewinder. In each simulation, a $10^5$ particle NFW-profile satellite was inserted at the apogalacticon of its orbit in a static, multi-component galaxy with a triaxial host halo, described below. The satellite was evolved first in isolation, then the host potential was turned on slowly over 10 satellite internal dynamical times to reduce artificial gravitational shocking. The initial position and velocity was obtained by integrating the orbit of the Sagittarius dwarf galaxy from present-day coordinates ${\bf r}  =(19.0, 2.7, -6.9)\ \mathrm{kpc}, \ {\bf v} = (230., -35., 195.) \ \mathrm{km\ s^{-1}}$ \citep{law10} backwards for $\sim$6 Gyr. The N-body satellite was then reintegrated from that initial position for the same interaction time and number of time-steps. This orbit has its apogalacticon at approximately 59 kpc and perigalacticon near 12 kpc with an average orbital period of 930 Myr, although all of these quantities vary over the course of the simulation due to the halo's triaxiality. Total energy is conserved to $\sim$2\% of the satellite internal potential energy.

To capture the character of streams across a range of merger mass ratios, four satellites with masses $\mathrm{m} = 2.5 \times 10^6,\ 2.5 \times 10^7,\ 2.5 \times 10^8,\ 2.5 \times 10^9\ \msun$, where m is the mass enclosed within 35 NFW scale radii, were evolved in the setup described above. Their scale radii, $r_0$, were adjusted to maintain a constant density across all simulations which results in identical fractional mass loss rates: 74\% of the initial mass is lost by the end of each simulation. The base value is $\mathrm{r_0}=0.01565$ kpc at $\mathrm{m} = 2.5 \times 10^6\ \msun$.

We take care to record the time at which each particle is unbound from the satellite in the simulations: we locate the position of the remnant iteratively by first calculating the satellite potential with all particles, removing particles with kinetic energies sufficient to escape, then recalculate the potential without those particles until the system converges. There is no spatial restriction on where a particle may become unbound and all particles (both bound and unbound) contribute their gravity during normal time-steps, presuming that they are resolved by the basis functions. 

In all simulations, the host potential is taken to be a three-component sum of a Miyamoto-Nagai disk \citep{miyamoto75}, Hernquist spheroid, and a triaxial, logarithmic halo \citep[e.g.,][]{law10}:
\begin{align}
	&\Phi_{\rm disk}(R,z) = -\frac{GM_{\rm disk}}{\sqrt{R^2 + (a + \sqrt{z^2 + b^2})^2}}\\
	&\Phi_{\rm spher}(r) = -\frac{GM_{\rm spher}}{r + c}\\
	&\Phi_{\rm halo}(x,y,z) = \vhalo^2 \ln(C_1 x^2 + C_2 y^2 + C_3 xy + (z/q_z)^2 + r_h^2)
\end{align}
where $C_1$, $C_2$, and $C_3$ are combinations of the $x$ and $y$ axis
ratios ($q_1$, $q_2$) and orientation of the halo with respect to the
baryonic disk ($\phi$):
\begin{align}
  C_1 &= \frac{\cos^2\phi}{q_1^2} + \frac{\sin^2\phi}{q_2^2}\\
  C_2 &= \frac{\sin^2\phi}{q_1^2} + \frac{\cos^2\phi}{q_2^2}\\
  C_3 &= 2\sin\phi\cos\phi \left(q_1^{-2} - q_2^{-2}\right).
\end{align}
The total potential then is just
\begin{equation}
	\Phi_{\rm tot} = \Phi_{\rm disk} + \Phi_{\rm spher} + \Phi_{\rm halo}\label{eq:lm10}.
\end{equation}
\begin{figure*}[!ht]
\begin{center}
\includegraphics[width=\textwidth]{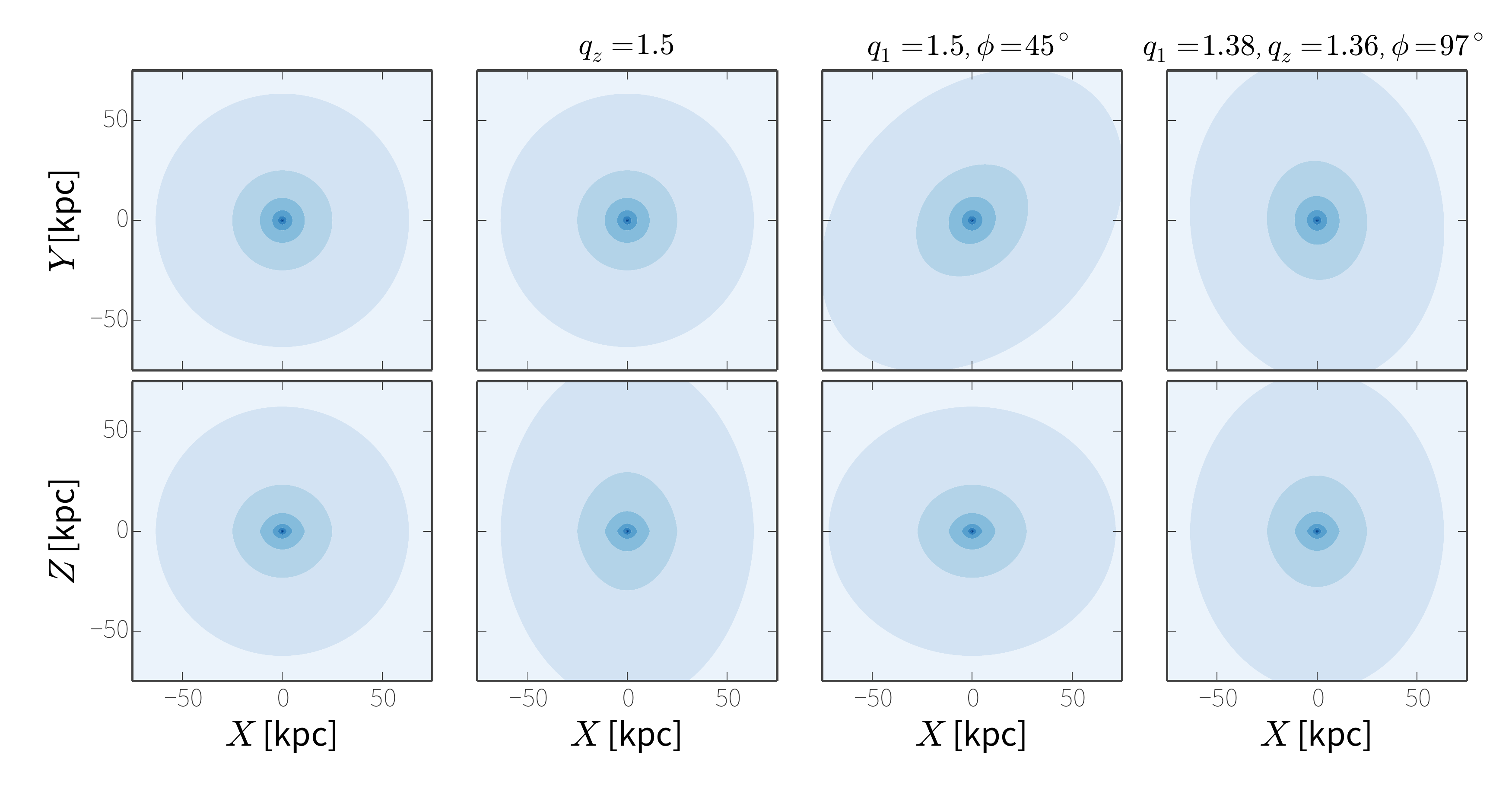}
\caption{ Equipotential contours for the LM10 potential (Eq.~\ref{eq:lm10}) in Galactocentric, cartesian coordinates for various halo parameter choices. For all panels, $\vhalo=121.858~\mathrm{km}/\mathrm{s}$, $r_h=12~\mathrm{kpc}$, and $q_2=1$. Left to right, each column represents a new choice of parameters. If not specified, other parameters are fixed to $q_1=q_z=1$ and $\phi=0^\circ$ (far left panels). Panels on far right show the best-fit parameter values from LM10. }\label{fig:potential}
\end{center}
\end{figure*}

We use this potential not because we think it is a realistic representation of the Galactic potential, but because successful inference with this potential demonstrates that it is possible to recover information about non-trivial potential forms. Figure~\ref{fig:potential} shows equipotential slices in the Galactic X-Z plane (Y=0) and Y-Z plane (X=0) for a few choices of  $q_1$,  $q_z$, and $\phi$ while holding all other parameters fixed, as described in the figure caption. The potential parameters used for the simulations are shown in the ``Truth'' column of Table~\ref{tbl:params}.

\begin{figure*}[!ht]
\begin{center}
\includegraphics[width=\textwidth]{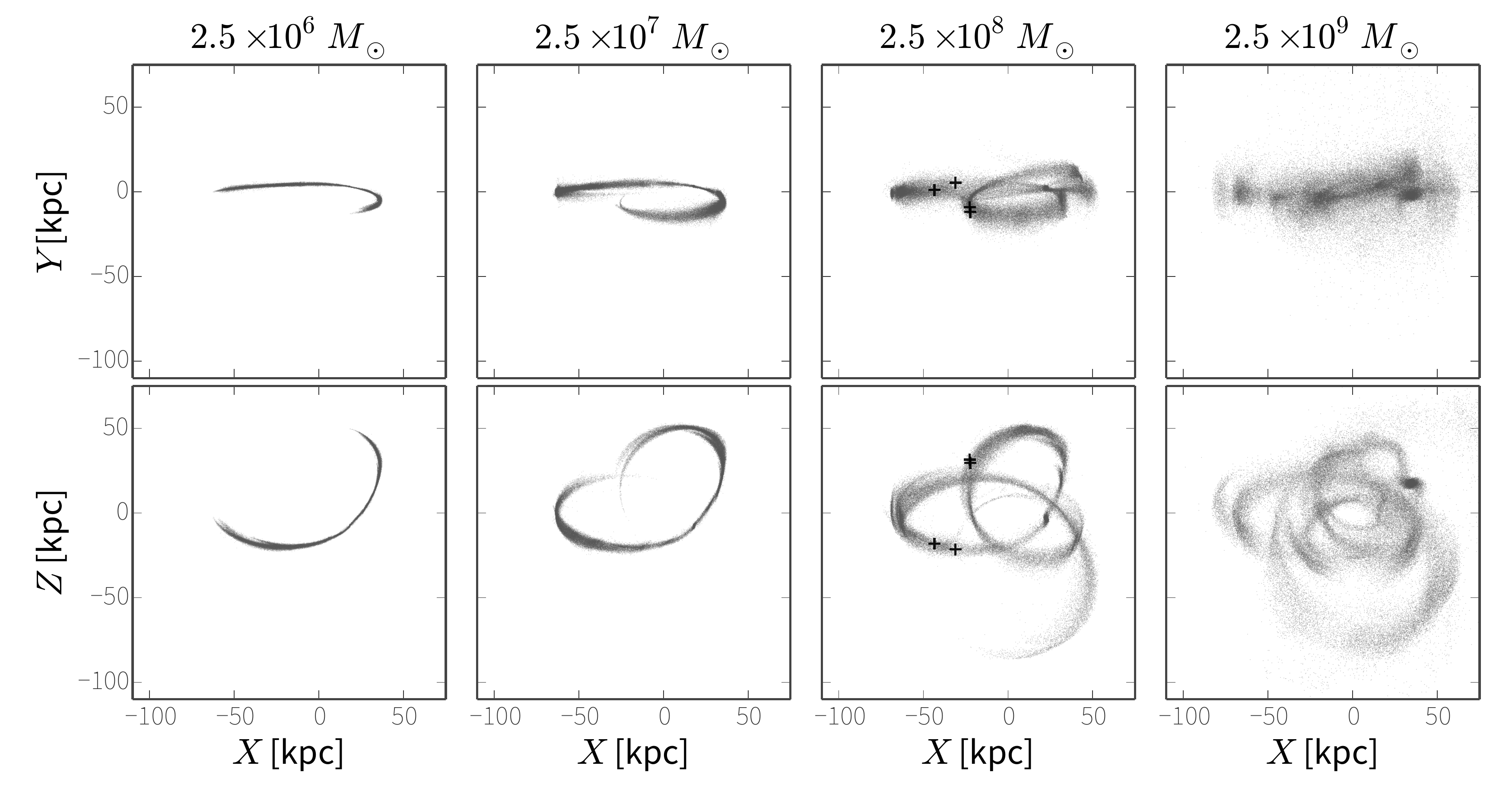}
\caption{ Particle positions (grey dots) in Galactocentric cartesian coordinates from the final time-step of four N-body simulations (Section~\ref{sec:sims}) with the same progenitor orbit initial conditions over a range of progenitor masses (columns). Black crosses indicate four particles chosen from each mass simulation and used in the experiments described in Section~\ref{sec:experiments}.}\label{fig:sims}
\end{center}
\end{figure*}

\section{Method (\rewinder)}\label{sec:method}

\rewinder\ integrates stars and the progenitor back from their present-day, observed positions to the time at which they become unbound from the satellite, where we evaluate the likelihood for each star. Below we (1) motivate a simple model for the debris just as it disrupts and (2) present this idea within a probabilistic framework for incorporating observational uncertainties and missing data. For a satellite galaxy orbiting within a static host potential, mass loss is driven by a combination of tidal stripping by the steady tidal field of the parent system and tidal shocking by the rapidly changing tidal field as the progenitor system moves through pericenter \citep[e.g.,][]{choi09}. On a mildly eccentric orbit, the tidal shocking is subdominant to the steady disruption. The interplay between these two processes as a function of orbital properties is outside the scope of this paper and will be explored in future work; we focus here on the slow removal of stars by steady tidal forcing. 

\subsection{Motivation: tidal debris}\label{sec:debris}

Consider a point-mass satellite with mass $m$ on a circular orbit with frequency $\Omega$ around a more massive ``host'' mass $M\gg m$ \citep[the ``restricted three-body problem''; e.g., \S 8.3][]{binneytremaine}. In a frame rotating with the orbital frequency of the satellite, the satellite remains fixed and the static, effective potential around the satellite has (amongst others) two unstable optima located around galactocentric radii $\sim R \pm r_J$, where $R$ is the orbital radius of the satellite and $r_J$ is the Jacobi or tidal radius
\begin{equation}
	r_J \sim R\left(\frac{m}{3M}\right)^{1/3}.\label{eq:ptmass}
\end{equation}
Particles that would be bound to the satellite in isolation may have enough (Jacobi) energy to overcome the effective potential barrier at these Lagrange points (Fig.~8.6 of BT) and thus will be preferentially stripped from the satellite at these points. For a spherical, extended parent mass distribution the tidal radius instead scales with the enclosed mass at the instantaneous orbital radius, $R(t)$, with additional terms that account for the local slope of the density profile. In general the satellite mass also depends on time. The tidal radius of the debris then follows
\begin{equation}
	r_{\rm tide}(t) = R(t)\left(\frac{m(t)}{3M_{\rm enc}(R)}\right)^{1/3}\label{eq:tidalradius}
\end{equation}
where $M_{\rm enc}$ is the instantaneous enclosed mass of the parent system within orbital radius $R$.\footnote{Note that this is the instantaneous orbital radius, \emph{not} the orbital radius at pericenter.}
In general, the Lagrange points may not be symmetric about the center of the satellite and may deviate from the tidal radius by factors of order unity. 

The bulk velocity of the satellite will be of order $V\sim \sqrt{GM_{\rm enc}/R}$. If the velocity dispersion of the satellite is $\sigma_v \sim \sqrt{Gm/r_J}$, then it follows
\begin{equation}
	\sigma_v(t) \sim V\left(\frac{m(t)}{M_{\rm enc}(t)}\right)^{1/3}\label{eq:velscale}.
\end{equation}
\citep[as pointed out in][]{binneytremaine} and hence we expect the debris-star velocities also to scale with $(m/M_{\rm enc})^{1/3}$. These scalings assume that the satellite is spherical with isotropic velocities; non-random, internal satellite orbits (e.g., a disk) will break these assumptions. 

In a triaxial potential, the orbital plane of the satellite is not fixed but we still expect there to be \emph{effective} Lagrange points for a spherical satellite system along the line of centers connecting the origin of the parent potential to the satellite --- that is, we expect the stars to be stripped at some characteristic distance from the satellite (near the tidal radius) along the instantaneous position vector of the satellite with some dispersion about these points. We proceed by defining a coordinate system relative to the position and velocity of the progenitor, ($\bs{r}_p$,$\bs{v}_p$), rotated into the instantaneous orbital plane defined by $\bs{L} = \bs{r}_p \times \bs{v}_p$. The (time-dependent) basis vectors are given by
\begin{align}
	\hat{\bs{x}}_1 &= \frac{\bs{r}_p}{\|\bs{r}_p\|}\label{eq:x1}\\
	\hat{\bs{x}}_2 &= \frac{\bs{L} \times \hat{\bs{x}}_1}{\|\bs{L} \times \hat{\bs{x}}_1\|}\\
	\hat{\bs{x}}_3 &= \hat{\bs{L}} = \frac{\bs{L}}{\|\bs{L}\|} = \frac{\bs{r}_p \times \bs{v}_p}{\|\bs{r}_p \times \bs{v}_p\|}\label{eq:x3}.
\end{align}
\begin{figure}[p]
\begin{center}
\includegraphics[width=\textwidth]{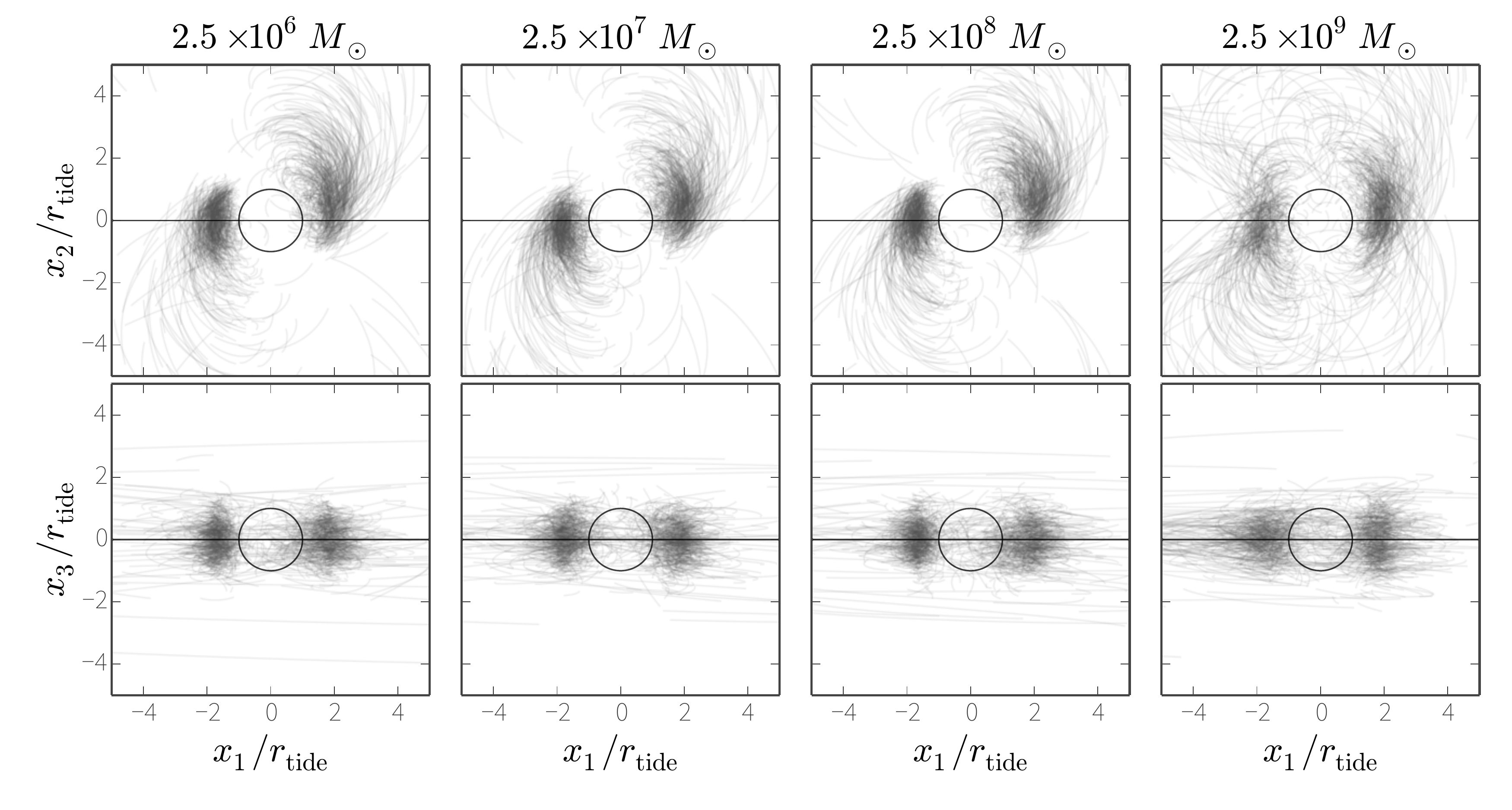}
\caption{ Orbits of 2000 randomly-drawn, disrupted particles projected into the instantaneous orbital plane coordinates (Eqs.~\ref{eq:x1}-\ref{eq:x3}), normalized by the tidal radius (Eq.~\ref{eq:tidalradius}), and only shown within half a satellite crossing time around $t=\tub$ for each of the four progenitor masses. The orbits were integrated backwards from their present-day positions (final time step of the N-body simulations) as test particles without the potential of the progenitor. Horizontal black line shows $x_2=0$ (top panels) and $x_3=0$ (bottom panels), and the unit circle (black circle) illustrates the classical disruption radius in these coordinates. }\label{fig:lpts_r}
\end{center}
\end{figure}
Figure~\ref{fig:lpts_r} shows sections of particle orbits (for particles that are stripped) from the N-body simulations described above, projected into this coordinate system. The positions are normalized by the instantaneous tidal radius (Eq.~\ref{eq:tidalradius}) and velocities are normalized by the instantaneous velocity scale (Eq.~\ref{eq:velscale}). Sections of the orbits symmetric around their unbinding times, $\tub$, are shown for each of the four progenitor masses. In these coordinates, the classical Lagrange points would be located at $x_1\approx\pm\rtide,x_2=0,x_3=0$ (illustrated by the intersection of the black unit circles and horizontal lines in Figure~\ref{fig:lpts_r}). Though not exactly centered on the point-mass Lagrange points derived for a circular orbit, the location of and dispersion about the effective Lagrange points in these scaled coordinates remain remarkably consistent across the range of progenitor masses explored. Figure~\ref{fig:lpts_v} shows the velocity orbits of each star also projected into these coordinates and normalized. The dispersion in velocity is well-normalized by the velocity scale of Eq.~\ref{eq:velscale}, though it is clear that the velocity dispersion along $\hat{x}_1$, the radial vector to the satellite position, is larger than that in other dimensions, possibly because of mild tidal shocking. 
\begin{figure}[ph]
\begin{center}
\includegraphics[width=\textwidth]{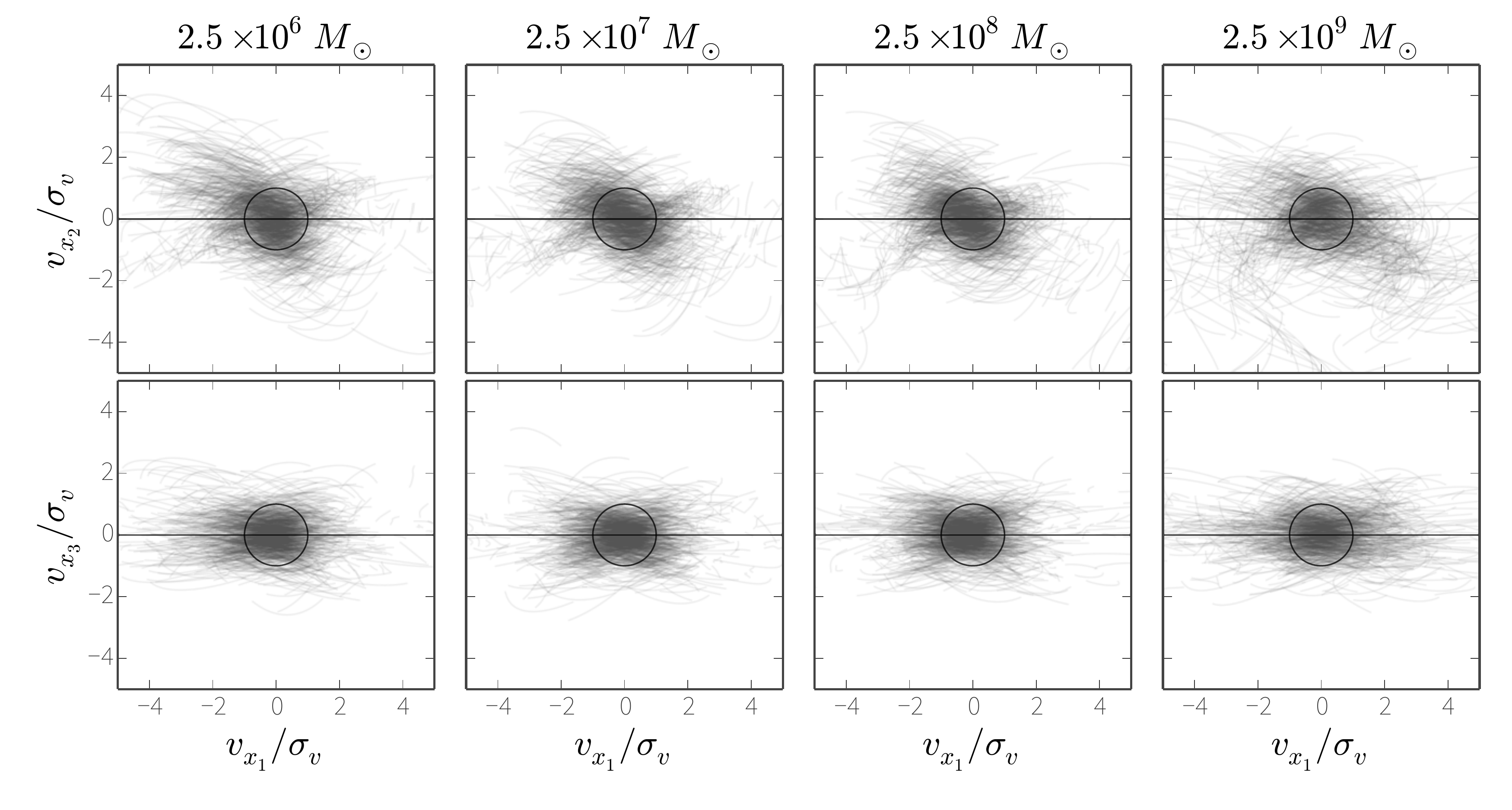}
\caption{ Same as Figure~\ref{fig:lpts_r} but for particle velocities normalized by the velocity scale (Eq.~\ref{eq:velscale}). }\label{fig:lpts_v}
\end{center}
\end{figure}

We conclude that even on a non-circular orbit in a complex potential, the dispersion --- in position and velocity --- of tidally stripped debris as it comes unbound from the satellite scales with the mass ratio $(m / M_{\rm enc})^{1/3}$. This motivates a model for tidal debris in which each star was ``released'' at the instantaneous, effective Lagrange point at its unbinding time, $\tub$, with a dispersion in position and velocity, all of which depend only on the mass and orbit of the progenitor and the parent potential. We present this model in detail below.

\subsection{Probabilistic model}
Suppose we observe the 6D position of a star, $\D = (l,b,d,\mu_l,\mu_b,v_r)$, in heliocentric coordinates --- e.g., the measured position on the sky, ($l$, $b$); distance, $d$; proper motions, ($\mu_l$, $\mu_b$); and line-of-sight velocity, $v_r$ --- and have determined through some other means that this star was once part of a progenitor system (e.g., satellite galaxy) with mass $m(t)$ that is disrupting and forming a cold debris structure in the potential, $\Phi$, of the parent galaxy (e.g., the Milky Way). We assume that the mass of the parent system enclosed within the pericenter of the orbit of the satellite, $M_{\rm enc}(R_{\rm peri})$, is much larger than the initial mass of the satellite, $M_{\rm enc}(R_{\rm peri})\gg m(t=0)$. The present position of the progenitor is observed to be at heliocentric position $\D_\sat$ (where any subscript $\sat$ refers to the progenitor). In general, the data for the star and progenitor will have significant uncertainties or missing dimensions at distances typical to the Galactic halo, thus we define $\W$ and $\W_\sat$ as the true 6D, heliocentric positions of the star and progenitor and will include these in our model. Since we include these positions as parameters, this method will work even when the star has missing dimensions or the progenitor location is unknown \citep[as in the Orphan stream,][]{belokurov07}.

To model the true, 6D, present-day position of the star, $\W$, we first transform to cartesian, Galactocentric coordinates where ($\bs{r}_0,\bs{v}_0$) and ($\bs{r}_{\sat,0},\bs{v}_{\sat,0}$) are the position and velocity of the star and progenitor today. The star is taken as having been sampled at $t=\tub$ (the unbinding time) from an isotropic Gaussian centered on one of the effective Lagrange points in position, and an isotropic Gaussian centered on the origin in velocity. The present-day phase-space position is the result of integrating this sample forward from $\tub$ in the parent potential, $\Phi$, whose form is parametrized by the vector $\bs{\Theta}_\Phi$. We assume that once the star becomes ``unbound'' from the satellite the potential of the satellite can be ignored,\footnote{This assumption breaks down for sufficiently large-mass progenitors, somewhere between $\sim10^8-10^9~\msun$.} and thus we treat the star as a test particle. The satellite mass enters this prescription through the tidal radius; a changing satellite mass will skew the positions of the effective Lagrange points and velocity scale. We allow for mass-loss by incorporating the initial satellite mass, $m_0$, and a constant mass-loss term, $\dot{m}$ into the model: $m(t) = m_0 - \dot{m}t$. Finally, we add two additional parameters: a constant, global factor, $\Loffset$, that scales the position of the effective Lagrange points relative to the classical tidal radius, and a binary parameter $\tailbit$ (one for each star), that is either $-1$ or $+1$ depending on whether the star is in the leading or trailing tail. To compress notation, we pack all progenitor parameters into the vector $\bs{\Theta}_\sat = (\W_\sat, m_0, \dot{m}, \Loffset)$. The likelihood for this model is:
\begin{align}
	p(\W \given \bs{\Theta}_\sat, \bs{\Theta}_\Phi, \tailbit, \tub) &= p(\bs{r}_0,\bs{v}_0, \given \bs{\Theta}_\sat, \bs{\Theta}_\Phi, \tailbit, \tub)\,\left\vert\J\right\vert\label{eq:starlike}\\
	p(\bs{r}_0,\bs{v}_0, \given \bs{\Theta}_\sat, \bs{\Theta}_\Phi,\tailbit,\tub) &= \left[\mathcal{N}(\bs{r} \given \bs{\Theta}_\sat, \tailbit)\,\mathcal{N}(\bs{v} \given \bs{\Theta}_\sat)\right]_{\tub}
\end{align}
where $\rtide$ is given by Eq.~\ref{eq:tidalradius}, and $\sigma_v$ by Eq.~\ref{eq:velscale} and the positions and velocities ($\bs{r},\bs{v}, \bs{r}_{\sat},\bs{v}_{\sat}$) are evaluated at the unbinding time, $\tub$, by integrating the orbits backwards from ($\bs{r}_0,\bs{v}_0, \bs{r}_{\sat,0},\bs{v}_{\sat,0}$). $\left\vert\J\right\vert = \left\vert\frac{\partial(x,y,z,v_x,v_y,v_z)}{\partial(l,b,d,\mu_l,\mu_b,v_r)}\right\vert$ is the absolute value of the determinant of the Jacobian that defines the transformation from heliocentric, spherical to Galactocentric, cartesian coordinates. We then marginalize the likelihood over all possible unbinding times, assuming a uniform prior of unbinding times over the entire interaction time:
\begin{align}
	p(\W \given \bs{\Theta}_\sat, \bs{\Theta}_\Phi, \tailbit) &= \int^{t_{\rm int}}_0 p(\W \given \bs{\Theta}_\sat, \bs{\Theta}_\Phi, \tailbit, \tub)\,p(\tub)\,d\tub
\end{align}
where the interaction time is taken to be $t_{\rm int}=6.2$~Gyr, the total duration of the simulations of Section~\ref{sec:sims}.

The likelihood above is evaluated for each star at all time-steps of the simulation during the marginalization with a uniform prior over all possible unbinding times (no assumptions are made about the orbital phase dependence of mass-loss). We chose this approach because the phase-space position (and hence time) at which a star becomes unbound from a satellite can only be strictly defined for the case of a spherical satellite orbiting on a circular orbit in a spherical potential. Various authors have chosen to measure mass loss in simulations of satellite disruption on eccentric orbits in non-spherical potentials by either looking at the kinetic energy of the particles relative to the satellite's potential energy \citep{johnston95} or looking at the point at which a star crosses the instantaneous tidal radius \citep{kuepper12}. Neither choice is strictly valid and both have been found useful for representing the mass-loss process. Hence we make the simplest choice of using Gaussians to characterize the phase-space distribution of debris as it makes the transition between being bound and unbound. We find that changing this distribution to a log-normal distribution does not affect our conclusions. 

We assume that the observational uncertainties are Gaussian in heliocentric, spherical coordinates, such that
\begin{align}
	p(\D \given \W) &= \mathcal{N}(\D \given \W, \bSigma)\label{eq:obsstar}\\
	p(\D_\sat \given \W_\sat) &= \mathcal{N}(\D_\sat \given \W_\sat, \bSigma_\sat)\label{eq:obsprog}
\end{align}
where $\mathcal{N}$ represents the normal distribution, and the covariance matrices $\bSigma$ and $\bSigma_\sat$ specify the observational uncertainties on the observed 6D position of the star and progenitor, respectively. 

For a single star, the evaluation of the likelihood works as follows: given observed, present-day coordinates for the star and progenitor ($\D,\D_\sat$), potential parameter values ($\bs{\Theta}_\Phi$), and nuisance parameter values ($\Loffset,\tailbit$),
\begin{enumerate}
	\item transform star and progenitor positions in heliocentric coordinates to Galactocentric, cartesian coordinates;
	\item integrate star and progenitor orbits backwards as test particles in the potential given by $\bs{\Theta}_\Phi$ for the total interaction time, $t_{\rm int}$ (in this case, $\sim$$6.2$~Gyr);
	\item transform the particle position to the relative, normalized coordinates defined in Section~\ref{sec:debris};
	\item compute the likelihood given by Equation~\ref{eq:starlike} at each time-step of the integration and marginalize over the unbinding time, $\tub$.
\end{enumerate}

Assuming each star is an independent tracer, the full likelihood for many stars is then just the product of the individual likelihoods:
\begin{equation}
	p(\{\D^{(i)}\}, \D_\sat \given \{\bs{\Theta}^{(i)}\}, \bs{\Theta}_\sat, \bs{\Theta}_\Phi ) = 
		p(\D_\sat \given \W_\sat) \, \prod_i \, p(\D^{(i)} \given \W^{(i)}) \,
			p(\W^{(i)} \given \bs{\Theta}_\sat, \bs{\Theta}_\Phi, \tailbit^{(i)})
\end{equation}
where $\bs{\Theta}^{(i)} = (\W^{(i)}, \tailbit^{(i)})$. All parameters are summarized in Table~\ref{tbl:params}.

\section{Experiments} \label{sec:experiments}
In what follows, we use \rewinder\ to model ``observations'' of a small sample of stars from one of the N-body simulations described in Section~\ref{sec:sims}. We consider this a strong test of the method because the data are generated with N-body simulations and not with anything resembling the likelihood function presented above. For all tests in this article, we use the same functional form for the potential as used in the simulations (Equation~\ref{eq:lm10}). When recovering the potential, we hold fixed the disk and spheroid parameters (see Table~\ref{tbl:params}), along with one of the halo parameters: the scale radius, $r_h$. The priors on the remaining halo parameters are taken to be uniform over a conservative domain of realistic values: for $\vhalo$, 100-200~km/s corresponds to a range in solar circular velocities from $\sim$210-250 km/s (holding other parameters fixed); the range in axis ratios allow for prolate, oblate, and generic triaxiality; and $\phi$ is restricted to $\pm45^\circ$ around the true simulation value, $\phi = 97^\circ$.

For all experiments below, we use a Markov Chain Monte Carlo (MCMC) algorithm to sample from the posterior probability distribution given by our model. Standard MCMC algorithms (e.g., Metropolis-Hastings) update a single chain while exploring parameter space. We instead use an affine-invariant ``ensemble'' sampler \citep{goodman10} that, each step in parameter space, updates the positions of many ``walkers'' (the ensemble). This algorithm is implemented in the \texttt{Python} programming language \citep{foremanmackey13} and runs naturally in a parallelized environment such as the message passing interface (MPI). In each experiment we run the walkers for a large number of steps from starting positions described below, then throw away these samples and take the positions from the final step of this ``burn-in'' phase as a starting point for the samples used for inference. We compute the autocorrelation time for each sampled parameter (using \texttt{ACOR}\footnote{\url{http://www.math.nyu.edu/faculty/goodman/software/acor/}}$^{,}$\footnote{\url{https://github.com/dfm/acor}}) and thin the chains by taking every $\mathrm{max}(t_{\rm acor})$ sample to ensure the samples are close to independent.

\begin{table*}[ht]
\begin{center}
	\begin{tabular}{l c c l} \toprule
		\multicolumn{4}{l}{{\bf \emph{Milky Way parameters} ($\bs{\Theta}_\Phi$)}} \\
		\toprule
		Component & Parameter & Truth & Prior \\\toprule
		disk & $M_{\rm disk}$ & $1.0\times10^{11}\,\msun$ & fixed \\ 
		& $a$ & 6.5 kpc & fixed\\
		& $b$ & 0.26 kpc & fixed\\
		\midrule
		spheroid & $M_{\rm spher}$ & $3.4\times10^{10}\,\msun$ & fixed\\ 
		& $c$ & 0.7 kpc & fixed\\
		\midrule
		halo & $\vhalo$ & 121.858 & $\mathcal{U}(100,200)$ km/s \\
		& $q_1$ & 1.38 & $\mathcal{U}(1,2)$\\
		& $q_2$ & 1.0 & fixed\\
		& $q_z$ & 1.36 & $\mathcal{U}(1,2)$\\
		& $\phi$ & 97 & $\mathcal{U}(52,142)$ deg\\
		& $r_h$ & 12 kpc & fixed\\
		\toprule
		\multicolumn{4}{l}{{\bf \emph{Progenitor parameters} ($\bs{\Theta}_\sat$)}} \\
		\toprule
		position & $\bs{r}_{\sat,0}$ & -- & $\|\bs{r}_{\sat,0}\|\sim\mathcal{U}(0,200)$~kpc \\
		velocity & $\bs{v}_{\sat,0}$ & -- & $\|\bs{v}_{\sat,0}\|\sim\mathcal{U}(0,500)$~km/s\\
		Lagrange pt. offset & $\alpha$ & -- & $\mathcal{U}(0.5, 2.5)$\\
		initial mass & $m_0$ & $2.5\times10^8\,\msun$ & fixed\\
		mass loss & $\dot{m}$ & -- & $3.2\times10^4\,\msun$/Myr (fixed)\\
		\toprule
		\multicolumn{4}{l}{{\bf \emph{Star parameters} ($\bs{\Theta}^{(i)}$)}} \\
		\toprule
		position & $\bs{r}_0$ & -- & $\|\bs{r}_0\|\sim\mathcal{U}(0,200)$~kpc \\
		velocity & $\bs{v}_0$ & -- & $\|\bs{v}_0\|\sim\mathcal{U}(0,500)$~km/s\\
		tail assignment & $\beta$ & -- & $\pm1$~equally likely (fixed at truth)\\ 
		\bottomrule
		\end{tabular}
	\caption{Parameter values used in the experiments of Section~\ref{sec:experiments}. $\mathcal{N}$ is the normal (Gaussian) distribution, and $\mathcal{U}$ the uniform distribution. There are 11 parameters for the Milky Way potential, but only four are left free to vary; some parameters are fixed (denoted by ``(fixed)'') at the true values used in the N-body simulations that generated the fake test data. The progenitor has nine parameters --- the position, $\bs{r}_{\rm p}$, and velocity, $\bs{v}_{\rm p}$, vectors each contain three components --- but only five are left free to vary. The sky coordinates (e.g., Galactic $l$, $b$) are assumed to be known with negligible uncertainty. Each star has eight associated parameters, five of which are allowed to vary. Sky coordinates are fixed, along with the tail assignment (whether the star belongs to the leading or trailing tail). For inference with four stars, there are $4+5+4\times4=25$ free parameters. \label{tbl:params}}
\end{center}
\end{table*}

\subsection{Data with negligible uncertainties}\label{sec:exp1}

We test \rewinder\ using only four stars (particles) --- two from the leading tail, and two from the trailing tail --- randomly sampled from the $2.5\times10^8\,\msun$ simulation (Section~\ref{sec:sims}), assuming the observed 6D positions for both the stars and the progenitor have negligible uncertainties. The stars were required to have been stripped after the first pericentric passage and have a present-day distance within $40$ kpc of the Sun. Figure~\ref{fig:sims} (second column from right) shows the randomly chosen stars (black crosses) in Galactic coordinates, over-plotted on all other simulated particles (grey points). 

\begin{figure}[p]
\begin{center}
\includegraphics[width=0.5\textwidth]{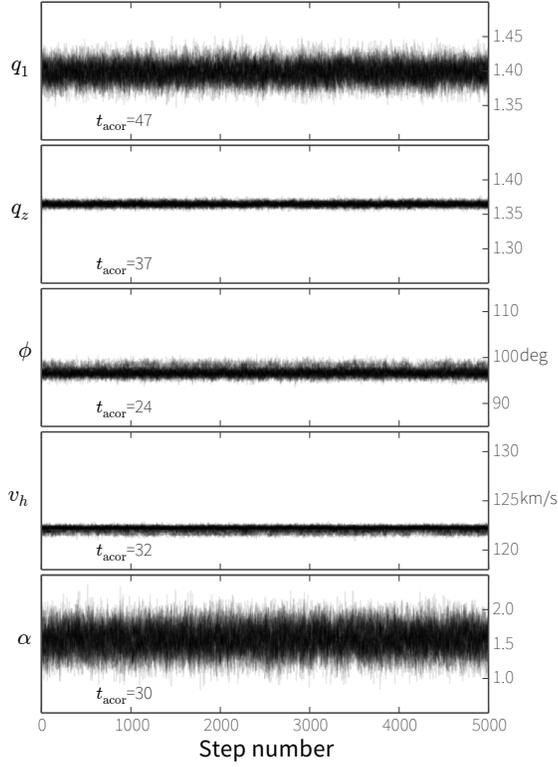}
\caption{ Positions of all 64 MCMC walkers (see Section~\ref{sec:experiments}) for each parameter at each step of the inference (black lines). The MCMC chains do not display any bulk movement or stray walkers, implying, by eye, that the chains have converged. We also compute the autocorrelation functions for each parameter and find that, for this experiment, the autocorrelation times are short (as shown on each panel in this figure). Projections of the binned samples (posterior densities) are shown in Figure~\ref{fig:exp1_posterior}. }\label{fig:trace}
\end{center}
\end{figure}

\begin{figure*}[!h]
\begin{center}
\includegraphics[width=\textwidth]{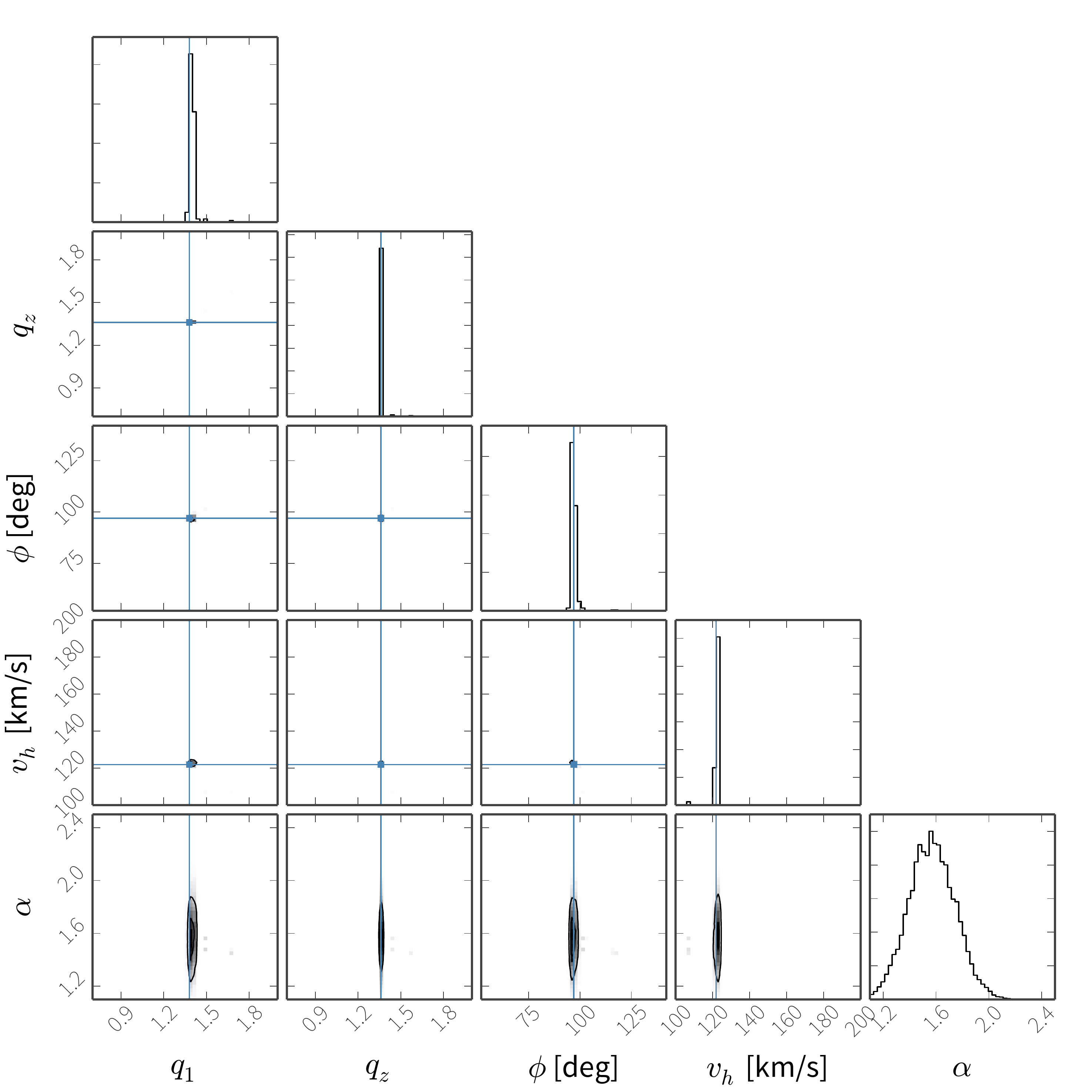}
\caption{ Projections of the posterior probability distribution over the four potential parameters (\potp) and Lagrange point offset ($\alpha$) assuming negligible uncertainties on the observed phase-space coordinates of eight stars and the progenitor, visualized as two-dimensional histograms of MCMC samples. Solid contours (black lines) show approximately 1$\sigma$ and 2$\sigma$ levels of the distribution. Vertical and horizontal lines (blue) show the true, input values for the potential parameters used in the N-body simulations. For the potential parameters, the axis ranges are chosen to be the same for this and the potential posterior plots to follow. }\label{fig:exp1_posterior}
\end{center}
\end{figure*}

We leave the potential parameters ($q_1,q_z,\phi,\vhalo$) free to vary, along with the Lagrange point offset, $\Loffset$, and initialize an ensemble of 64 MCMC walkers by sampling from the priors summarized in Table~\ref{tbl:params}. We run the walkers for 5000 steps to burn-in, then restart the sampler starting from the final position of the burn-in phase and run for another 5000 inference steps. Figure~\ref{fig:trace} shows the walker positions over the 5000 inference steps for each of the five parameters. The autocorrelation time for each parameter is displayed on its corresponding panel. Figure~\ref{fig:exp1_posterior} shows projections of the posterior probability distribution for the parameters; with perfect data, the uncertainties on the potential parameters are all $<1\%$.

\subsection{Data with near-future uncertainties}\label{sec:exp2}
We next take the same four stars used in the previous experiment and ``observe'' them with optimistic observational uncertainties. We require these stars to be RR Lyrae variables which are known to be excellent distance indicators via the mid-infrared period-luminosity relation \citep[as shown in, e.g.,][]{madore12}. Nearly 100 RR Lyrae associated with the Sgr stream and $\sim$30 associated with the Orphan stream will be observed with \spitzer as part of the SMASH survey \citep{smashprop} with expected fractional distance uncertainties around $\sim$2\%. These stars will also be included in the \gaia\, proper motion catalog: at a distance of $\sim$50~kpc, a typical RR Lyrae will have a tangential velocity error around $\sim$20~km/s, though our sample of test stars are all within 40~kpc.

For this experiment, we (1) assume the stars are RR Lyrae stars (e.g., bright distance indicators) such that the fractional distance uncertainty is $2\%$; (2) neglect the uncertainty in angular position, $l,b$ (for a typical RR Lyrae at 50~kpc this is $\sim$$10^{-7}$~deg for \gaia); (3) assume we can measure radial velocities to these stars with 5~km/s uncertainty; and (4) compute the sky-averaged \gaia\, proper-motion uncertainty for each star assuming an F0V spectral type using the \texttt{PyGaia}\footnote{\url{https://github.com/agabrown/PyGaia}} code and use this uncertainty for both components of proper motion. We further assume that we know the tail assignment for each star, $\beta$. We observe the position of the progenitor with the same observational uncertainties though in reality some coordinates will have even better measurements.

\begin{figure*}[!h]
\begin{center}
\includegraphics[width=\textwidth]{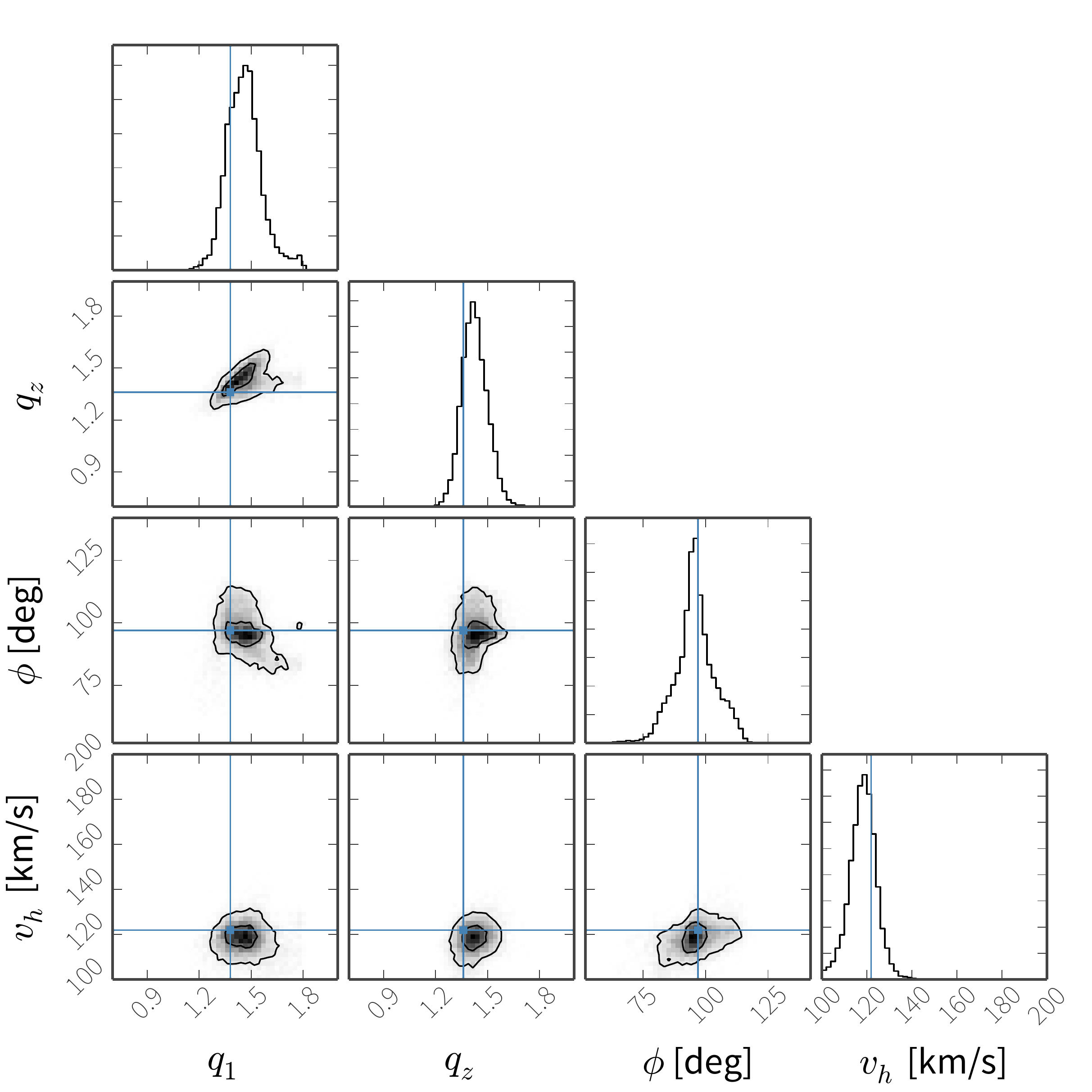}
\caption{ Projections of the marginal posterior over the triaxial potential parameters for observed stars and progenitor with near-future uncertainties (Section~\ref{sec:exp2}). Axis ranges show the lower and upper bounds on the uniform priors over these parameters. }\label{fig:exp2_potential}
\end{center}
\end{figure*}

\begin{figure*}[!h]
\begin{center}
\includegraphics[width=\textwidth]{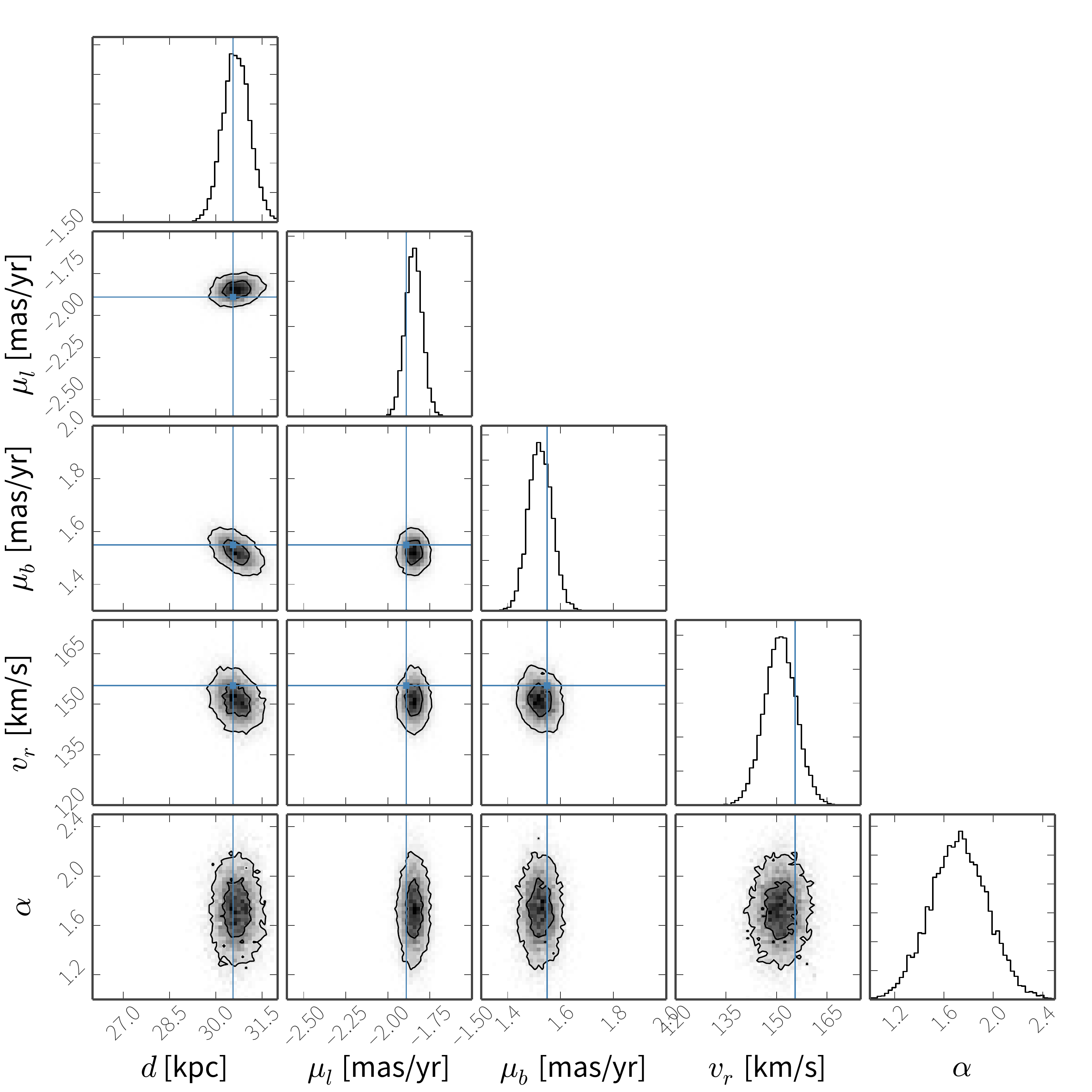}
\caption{ Projections of the marginal posterior over the progenitor parameters for observed stars and progenitor with near-future uncertainties (Section~\ref{sec:exp2}). }\label{fig:exp2_satellite}
\end{center}
\end{figure*}

\begin{figure*}[!h]
\begin{center}
\includegraphics[width=\textwidth]{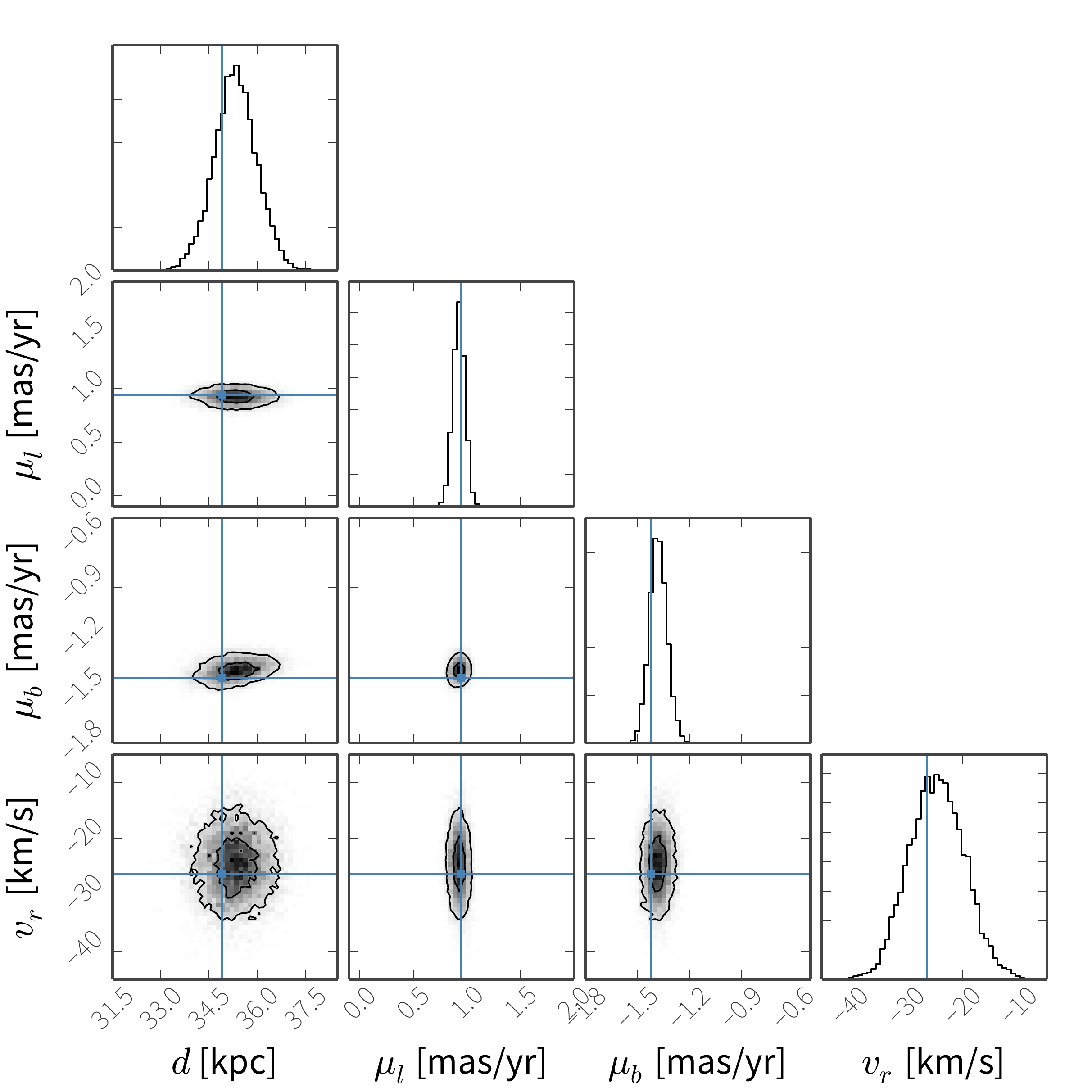}
\caption{ Projections of the marginal posterior over parameters for one of the stars for observed stars and progenitor with near-future uncertainties (Section~\ref{sec:exp2}).   }\label{fig:exp2_particle}
\end{center}
\end{figure*}

This experiment samples over the four potential parameters, the Lagrange point offset, four phase-space coordinate parameters for each star (16 total), and four phase-space coordinate parameters for the progenitor --- 25 parameters in total. We use an ensemble of 256 walkers and draw initial conditions for the coordinate parameters by sampling from Gaussian's centered on the ``observed'' value with variances specified by the observational uncertainties. Other parameters --- potential parameters and $\Loffset$ --- are initialized by drawing from the priors summarized in Table~\ref{tbl:params}. We burn in the walkers for 50000 steps and run for 100000 inference steps, but thin the chains by taking every $\max(t_{\rm acor})$ sample, where $\max(t_{\rm acor}) \approx 4000$~steps. Figures~\ref{fig:exp2_potential}, \ref{fig:exp2_satellite}, \ref{fig:exp2_particle} show the marginalized posteriors for the potential parameters, satellite parameters, and parameters for a single particle. The uncertainties on the potential parameters are only 5-7\%

\subsection{Precise distance measurements with missing proper motions}\label{sec:exp3}
The SMASH survey \citep{smashprop} will be completed within a year (early 2015), long before the final \gaia\, data release. Thus, we will soon have precise distance measurements for stars in the Sgr and Orphan streams, but poor or no proper motion constraints. SMASH also targets several RR Lyrae in the Sgr core, which will enable a high-precision distance measurement of the progenitor. Though proper motions have been measured for the Sgr progenitor, we now consider a case in which we have high-precision (2\%) distances to stream stars and the progenitor, 10~km/s radial velocity uncertainty, but missing proper motions for all stars and progenitor. 

\begin{figure*}[!h]
\begin{center}
\includegraphics[width=\textwidth]{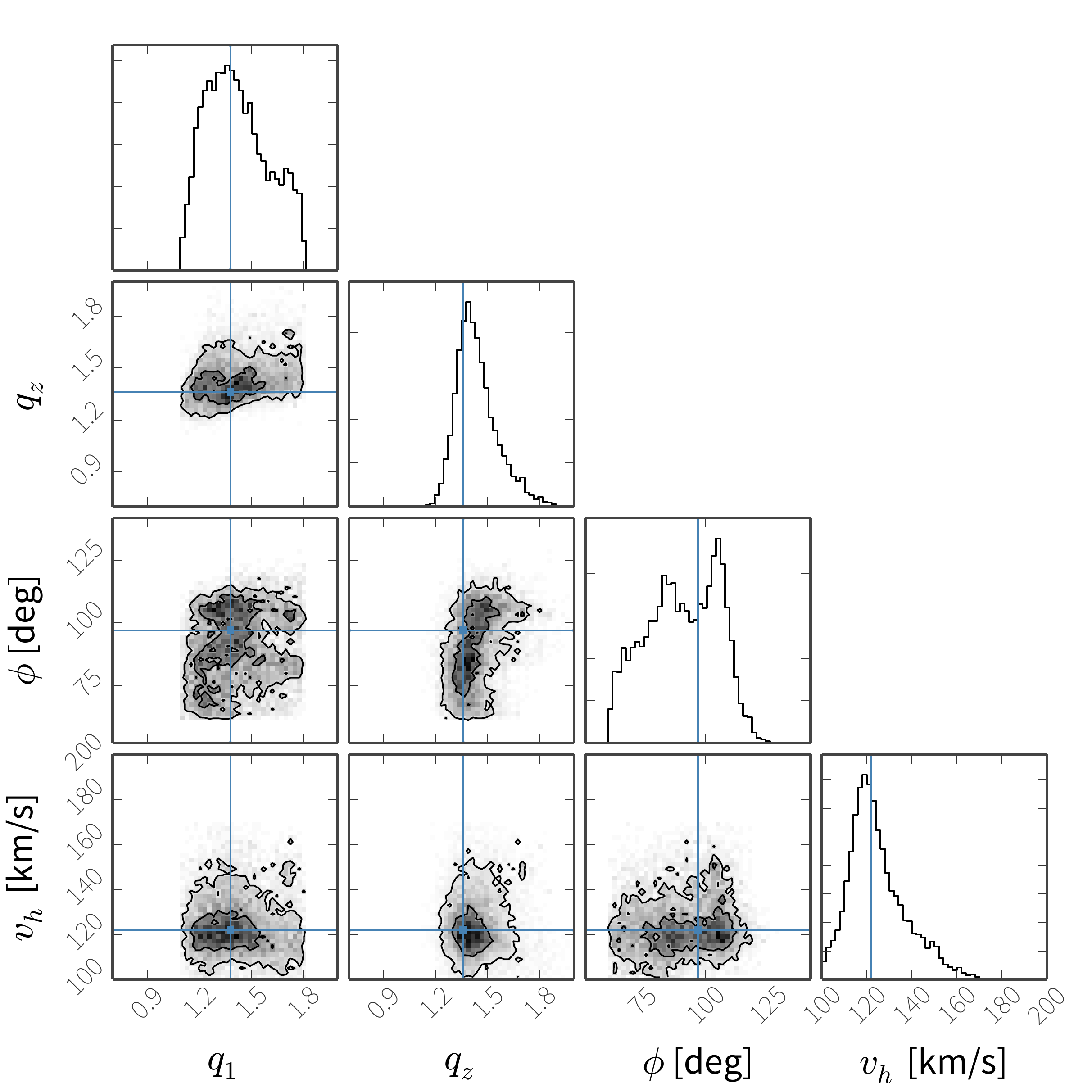}
\caption{ Same as Figure~\ref{fig:exp2_potential} but for data with no proper motion measurements. }\label{fig:exp3_potential}
\end{center}
\end{figure*}

\begin{figure*}[!h]
\begin{center}
\includegraphics[width=\textwidth]{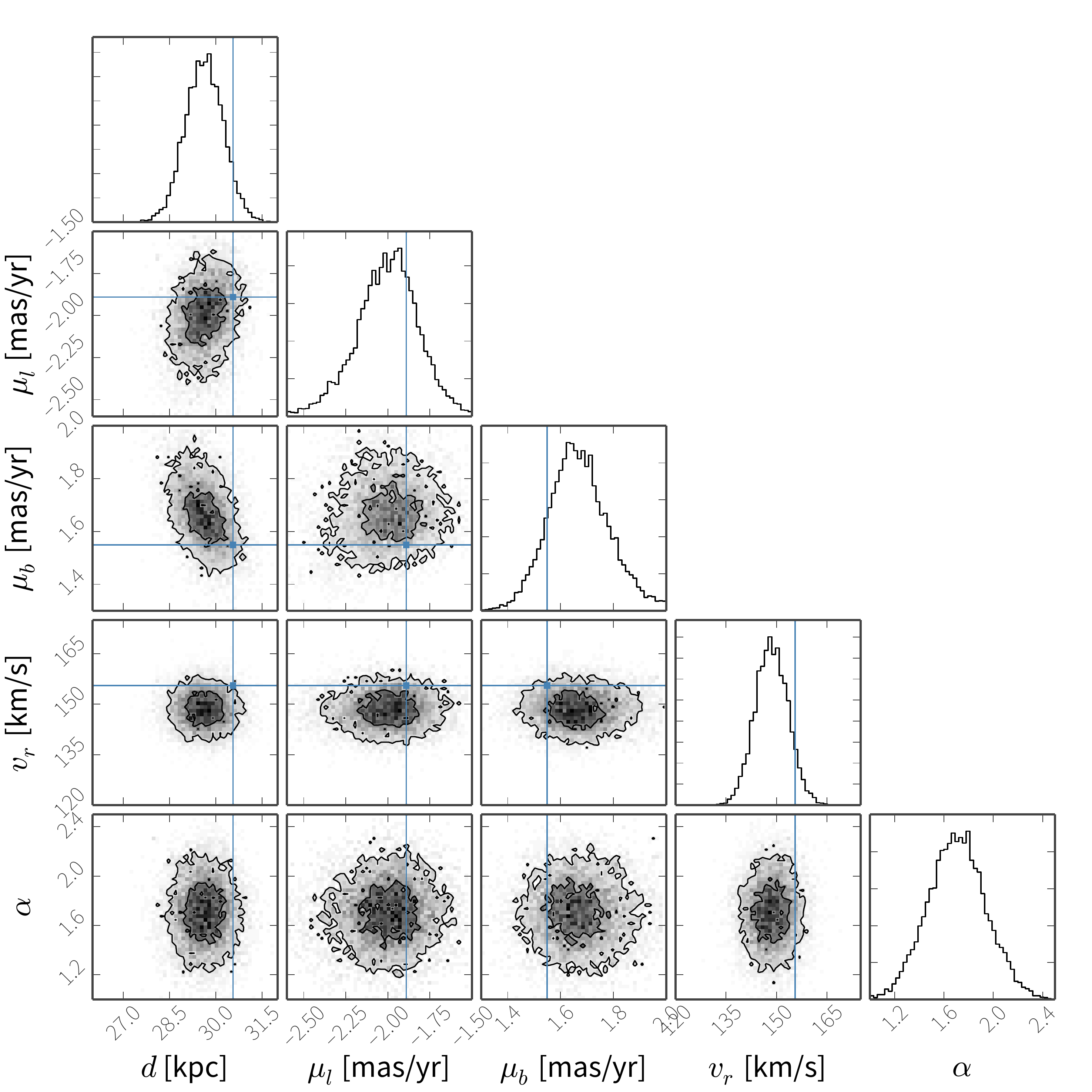}
\caption{ Same as Figure~\ref{fig:exp2_satellite} but for data with no proper motion measurements.  }\label{fig:exp3_satellite}
\end{center}
\end{figure*}

\begin{figure*}[!h]
\begin{center}
\includegraphics[width=\textwidth]{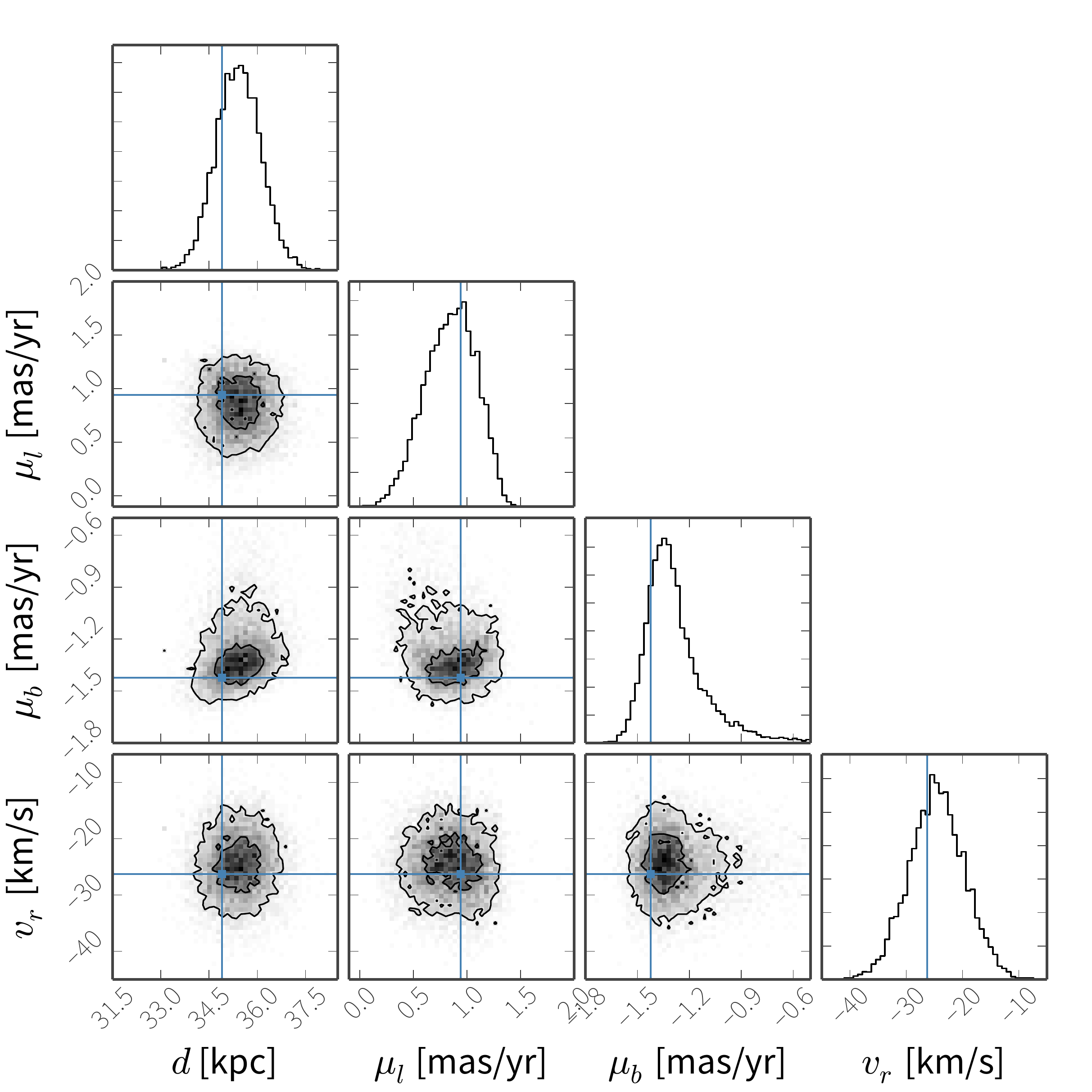}
\caption{ Same as Figure~\ref{fig:exp2_particle} but for data with no proper motion measurements.  }\label{fig:exp3_particle}
\end{center}
\end{figure*}

As with the previous experiment, this experiment includes 25 model parameters in total. We again use an ensemble of 256 walkers. Other parameters --- potential parameters and $\Loffset$ --- are initialized by drawing from the priors summarized in Table~\ref{tbl:params}. We burn in the walkers for 50000 steps and run for 500000 inference steps. We again thin the chains by taking every $\max(t_{\rm acor})$ sample, but here the autocorrelation time is found to be very long, $\max(t_{\rm acor}) \approx 15000$~steps. Figures~\ref{fig:exp3_potential}, \ref{fig:exp3_satellite}, \ref{fig:exp3_particle} show the marginalized posteriors for the potential parameters, satellite parameters, and parameters for a single particle. Even a small sample of stars with precise distance measurements provide an enormous amount of information about the triaxiality of the potential. Fractional uncertainties on the potential parameters are $\sim$15\% and the initially missing proper-motions are recovered with $\sim$25\% uncertainties.

\section{Discussion}\label{sec:discussion}

{\bf 1. Observational uncertainties:} Any method that uses tidal debris as a potential measure must also model the (significant) uncertainties on the kinematic measurements; \rewinder\ handles observational uncertainties and missing data dimensions by including the true 6D positions of stars and the progenitor as model parameters. In this article we have tested \rewinder\ using data of varied quality for a small sample of stars. 

{\bf 2. Form of the potential:} No assumptions were made about the the form of the Galactic potential in deriving the likelihood for \rewinder. The simple experiments in this article infer potential parameters from the same static, analytic potential used in the N-body simulations that generated the fake data. However, in principle the potential could be time-dependent, clumpy, or have properties that vary with radius. The orbits of the stars and progenitor are directly integrated and only require a function that evaluates the acceleration due to the potential at a given position (and time). We have shown that with the correct form for the Galactic potential, \emph{precise} measurements of the potential parameters are attainable with near-future data, but we have not discussed the \emph{accuracy} of such measurements due to incorrect assumptions. For example, Bonaca et al. (in prep.) show that using a static potential model to fit the live potential of the \project{Via Lactea} halo with tidal streams can introduce significant biases to measurements of the halo mass, even with perfect knowledge of the 6D coordinates of stars in the streams. In future work, we will (1) try fitting incorrect potential models to see if we can still recover global properties (e.g., flattening); (2) run simulations with smoothly changing potentials \citep[e.g.,][]{buist14} and attempt to model the time dependence; and (3) explore using a generic, non-parametric potential form (e.g., a basis function expansion) as the recovery potential model. 

{\bf 3. Multiple debris structures:} Each stream will provide constraints on different properties of the potential. For example, more eccentric streams may better constrain the radial profile \citep[see][who illustrate the power of using multiple streams to simultaneously constrain the potential using orbit fitting]{deg14}. In this article, we only consider a stream on a Sgr-like orbit, which lies nearly in the Galactic $x$-$z$ plane. We find that the stream puts better constraints on the $z$-axis flattening, $q_z$, than on the flattening in the $x$-$y$ plane, $q_1$, as one might expect. The best measurements of the detailed shape of the Galactic potential will come from combining constraints from multiple streams. We have defined above a proper likelihood function for observed stars in a given stream, thus incorporating multiple streams into the inference only requires multiplying the likelihoods computed for each stream and progenitor pair.

{\bf 4. Comparing models to data:} With \rewinder, each star adds constraints on the potential parameters and does not require matching a simulated density to the observed density of stars, thus this method works well even for small samples of well-measured stream stars. It might be that the most relevant stellar samples for inferring the Milky Way potential are small. For instance, stars that produce good distance estimates might be much more valuable than typical stars in the sample so that we may limit to only variable stars, e.g., RR Lyraes. These valuable stars are rare (and their abundances are age and metallicity dependent); there could be many cold structures in the Milky Way halo that are highly constraining on the potential in principle, but which contain only a few good distance-indicating members. It is not yet known what the trade-offs are between having many stars at low precision and a few at high precision, nor is it known how valuable distance information really is, when a structure contains many precisely observed members.

{\bf 5. Computational expense:} One point of concern for \rewinder\ is that the number of parameters scales with eight times the number of stars: each star has six coordinate parameters (the true position), the unbinding time, and the tail assignment. Computational constraints limit the sample size, but even so, the inference is much faster than full N-body modeling because the stars and progenitor are treated as test particles. Each step in parameter space with \rewinder\ requires integrating the orbits of stars and the progenitor: we presently integrate with a fixed time-step using leapfrog integration. Computing the acceleration due to the given potential at each step is implemented in \texttt{Cython} (and approaches \texttt{C}-like speed), but the rest of the code is written in pure-\texttt{Python}. Though already significantly less computationally intensive than running a full N-body simulation for every parameter step, running each MCMC walker for $\gtrsim$100000 steps requires parallelization and many hours of CPU time on a compute cluster. Further optimization of the integration method (e.g., using an adaptive method or implementing in \texttt{Cython}) could speed up the inference significantly. 

\section{Conclusions}\label{sec:conclusion}
We have presented a probabilistic model (\rewinder) --- a likelihood function and with priors on the parameters --- for using stars observed in tidal streams to constrain properties of any underlying gravitational potential. \rewinder\ relies on direct orbit integration and not on computing conserved quantities (e.g., actions) and can thus be used with arbitrarily complex (e.g., time-dependent or clumpy) forms of the potential. We have performed several experiments to show that \rewinder\ simultaneously constrains the potential and models a tidal stream given simulated data with a range of realistic uncertainties. We find that with future high-quality data --- that is, high precision distance measurements from RR Lyrae and proper motions from \gaia --- a small sample of just four stars in a tidal stream modeled with \rewinder\, provide measurements of potential parameters that rival present-day constraints from comparing full N-body simulations to large numbers of stellar tracers with poorly measured kinematics. For this high-quality data, we recover the input potential parameter values with uncertainties of order 5-7 percent. Without proper motion data the uncertainties are around 15 percent. We consider this work to be an encouraging first step towards the goal of recovering the (presumably) much more complex --- time-dependent, clumpy, with axis ratios and orientations that vary with radius --- Milky Way potential with larger data sets.

\acknowledgements
We thank Anthony Brown (Leiden) for insight on the \gaia\, data quality and for providing the open source \texttt{PyGaia} code for computing predicted uncertainties. We thank the organizers of the Gaia Challenge (2013) and the University of Surrey for hospitality during the workshop. We thank Ana Bonaca (Yale), Jo Bovy (IAS), Dan Foreman-Mackey (NYU), Marla Geha (Yale), Andreas K{\"u}pper (Columbia), and Hans-Walter Rix (MPIA) for useful discussions.
KVJ thanks the Institute of Astronomy at the University of Cambridge for hospitality and Vasily Belokurov (IoA) and his group for discussion of his work while there.
APW is supported by a National Science Foundation Graduate Research Fellowship under Grant No.\ 11-44155. This work was supported in part by the National Science Foundation under Grant No. PHYS-1066293 and AST-1312196. 
DWH was partially supported by the NSF (Grant IIS-1124794) and the Moore-Sloan Data Science Environment at NYU.
This research made use of Astropy, a community-developed core \texttt{Python} package for Astronomy \citep{astropy13}. This research has made use of NASA's Astrophysics Data System.
This work additionally relied on Columbia University's \emph{Hotfoot} and \emph{Yeti} compute clusters, and we acknowledge the Columbia HPC support staff for assistance. \\


\end{document}